\documentclass[useAMS,usenatbib]{mn2e}

\usepackage{times}  
\usepackage{listings}
\usepackage{graphics}
\usepackage{subfigure}
\usepackage{graphicx}
\usepackage{graphicx,xspace}
\usepackage{amssymb}
\usepackage{amsmath}
\usepackage{aas_macros}
\usepackage{lscape}
\usepackage{rotating}
\usepackage{placeins}
\usepackage{natbib}
\usepackage{color}
\usepackage{cuted}
     
\newcommand{\eagle}{{\sc{eagle}}\xspace}

\newcommand{\gadget}{{\sc{gadget-3}}\xspace}

\newcommand{\subfind}{{\sc{subfind}}\xspace}

\title[Baryonic mass budgets for haloes in \eagle]
{Baryonic mass budgets for haloes in the \eagle simulation, including ejected and prevented gas}

\author[P. D. Mitchell \& J. Schaye]{
\newauthor Peter D. Mitchell\thanks{\rm E-mail: mitchell@strw.leidenuniv.nl}$^{1}$ \&
Joop Schaye$^{1}$\\
$^{1}$Leiden Observatory, Leiden University, P.O. Box 9513, 2300 RA Leiden, the Netherlands\\
}

\begin{document}

\date{\today}
\pagerange{\pageref{firstpage}--\pageref{lastpage}} \pubyear{2021}
\maketitle
\label{firstpage}

\begin{abstract}
Feedback processes are expected to shape galaxy evolution 
by ejecting gas from galaxies and their associated dark matter 
haloes, and also by preventing diffuse gas from ever being accreted.
We present predictions from the \eagle simulation project
for the mass budgets associated with ``ejected'' and ``prevented''
gas, as well as for ejected metals. We find that most of
the baryons that are associated with haloes of mass
$10^{11} < M_{200} \, /\mathrm{M_\odot} < 10^{13}$ at $z=0$ have
been ejected beyond the virial radius after having been accreted.
When the gas ejected from satellites (and their progenitors) 
is accounted for,
the combined ejected mass represents half of
the total baryon budget even in the most massive
simulated galaxy clusters 
($M_{200} \approx 10^{14.5} \, \mathrm{M_\odot}$),
with the consequence that the total baryon budget exceeds
the cosmic average if ejected gas is
included. 
We find that gas is only prevented from being
accreted onto haloes for $M_{200} < 10^{12} \, \mathrm{M_\odot}$,
and that this component accounts for about half
the total baryon budget for $M_{200} < 10^{11} \, \mathrm{M_\odot}$,
with ejected gas making up most of the remaining half.
For metals, most of the mass that is not locked
into stars has been ejected beyond the virial radius,
at least for $M_{200} < 10^{13} \, \mathrm{M_\odot}$.
Finally, within the virial radius we find
that most of the mass in the circum-galactic
medium (CGM) has not passed through the ISM of 
a progenitor galaxy, for all halo masses and redshifts.
About half of the CGM within half the virial radius
has passed through the ISM in the past, however.
\end{abstract}

\begin{keywords}
galaxies: formation -- galaxies: evolution -- galaxies: haloes -- galaxies: stellar content
\end{keywords}

\section{Introduction}

Within the context of the working $\Lambda$ cold dark matter ($\Lambda$CDM) model, galaxy formation
is thought to trace the formation of dark matter haloes. Gas accretion is expected
to approximately trace dark matter accretion at the halo virial radius, and
thereafter the gas is thought to condense at the halo center by virtue of energy dissipation via
radiative cooling \cite[e.g.][]{White78}. Feedback processes then regulate the formation
of stars locally within the interstellar medium, eject gas back into
(and beyond) the halo, and modulate the infall of gas onto the interstellar medium (ISM)
(and possibly also onto the halo). Many authors have studied these effects
using cosmological zoom-in simulations 
\cite[e.g.,][]{Ubler14,Muratov15,Christensen16,AnglesAlcazar17},
as well as large-volume cosmological simulations \cite[e.g.,][]{Oppenheimer10,FaucherGiguere11,VanDeVoort11,Correa18b,Nelson19,Mitchell21}.

Recently, \cite{Mitchell20b,Mitchell20a} presented a complete set of measurements
of inflow and outflow rates of gas onto/from both haloes and galaxies in the \eagle simulations \cite[][]{Schaye15}.
Here, we supplement this information by presenting instead how the
masses in different galaxy/halo components are affected by these processes.
In particular, we address questions such as how many baryons have been
ejected from haloes, how much gas was prevented from being accreted
onto haloes in the first place, and how much of the circum-galactic medium (CGM)
has previously been part of the ISM of a galaxy. We acknowledge that these predictions are model
dependent and are not easily tested by observations; rather our intention
is to provide a pedagogical overview of how gas is accreted onto
and ejected from galaxies and haloes, providing a physically viable theoretical
scenario for how this might occur in reality. This in turn can be compared
to other state-of-the-art simulations in the future.

Some of these questions have already been addressed directly with other cosmological
simulations, or have been studied using simplified analytic modelling.
\cite{Hafen19} analyse the FIRE cosmological zoom-in simulations, quantifying
 how much of the CGM is comprised of gas that
was ejected from the ISM of galaxies, and how much of that
gas was ejected from the main-progenitor galaxy. They find that most
of the mass in the CGM has not been ejected from the ISM in the
past (and is labelled ``IGM accretion''). They also find that the origin of ISM-processed
CGM is dominated by the main progenitor galaxy for low-mass haloes
($M_{\mathrm{h}} < 10^{11} \, \mathrm{M_\odot}$), but that
the main progenitor (labelled ``Wind'') and satellite galaxies 
(labelled ``Satellite Wind'') provide comparable
contributions for $M_{\mathrm{h}} \sim 10^{12} \, \mathrm{M_\odot}$.
These findings are qualitatively consistent with what we find in \eagle,
although we find that the halo mass scale at which the ``satellite
wind'' component is important is at higher halo masses ($M_{200} \gtrsim 10^{13} \, \mathrm{M_\odot}$).
This is possibly because \eagle (unlike FIRE) includes AGN feedback that
drives outflows in $M_{200} \sim 10^{12} \, \mathrm{M_\odot}$ haloes,
increasing the relative importance of outflows from the main progenitor
in this mass range \cite[e.g.][]{Davies20,Oppenheimer20,Mitchell20a}. At this mass scale, \eagle also predicts significantly lower total
baryonic masses within the virial radius than FIRE, which we expect
for the same reason.

With a very different modelling approach, \cite{Afruni21} attempt to
reproduce observations of cool circum-galactic gas around local
star-forming galaxies using a simple model for how outflows are launched from
the ISM and then propagate through the CGM. They find that such
a model is not naturally capable of reproducing the observed
radial and velocity distributions of cool circum-galactic gas clouds,
and conclude that much of the cool CGM must have been accreted
from the IGM, without having been ejected from the central galaxy.
These conclusions are broadly consistent with
the measurements from \eagle that we present here.

The layout of this paper is follows. We provide details
of the simulations and analysis methodology in Section~\ref{method_sec},
our results are presented in Section~\ref{result_sec}, and we
summarise and present our conclusions in Section~\ref{summary_sec}.

\section{Methods}
\label{method_sec}

\subsection{Simulation}

Our analysis is performed on the \eagle project \cite[][]{Schaye15,Crain15}, which has been
publicly released \cite[][]{McAlpine16}. \eagle is a suite of cosmological simulations,
spanning a range of box sizes, resolutions, and model variations. \eagle uses a modified
version of the \gadget code \cite[last described in][]{Springel05b} to solve the equations of gravity
and hydrodynamics, employing smoothed particle hydrodynamics (SPH). A $\Lambda$CDM cosmological
model is assumed, with parameters set following \cite{Planck14}. Simple subgrid models
are included to model star formation and stellar evolution, supermassive black hole (SMBH) 
formation and evolution, feedback from stars and active galactic nuclei (AGN), 
and radiative cooling and heating.

We use exclusively simulations run with the Reference \eagle model \cite[see][for details]{Schaye15},
which was calibrated to be consistent with observed star formation thresholds and $\mathrm{kpc}$-scale
efficiencies, and to broadly reproduce the observationally inferred galaxy
stellar mass function, trends of galaxy size with stellar mass, and of SMBH mass with
galaxy stellar mass, all at $z \approx 0$. Most of our analysis is performed on the largest 
available $(100 \, \mathrm{Mpc})^3$-volume simulation, run with $1504^3$ particles
in both dark matter and gas, with a particle mass of 
$9.7 \times 10^6 \, \mathrm{M_\odot}$ for dark matter, initial gas
particle mass of $1.8 \times 10^6 \, \mathrm{M_\odot}$, and with maximum physical
gravitational softening of $0.7 \, \mathrm{kpc}$ (for both dark matter, gas, and also stars and black holes).

\subsection{Measurements}
\label{measure_sec}

Our analysis is based on following Lagrangian SPH gas particles across simulation outputs,
keeping track of which particles are accreted and then ejected from specific haloes and galaxies.
This information is passed forward in time from subhalo progenitors to descendants, merging
the lists of ejected particle identifiers when two subhaloes merge.
Accretion/ejection from haloes is defined via a simple spherical radial cut at the halo
virial radius, which we define as $R_{200}$, the radius enclosing a mean overdensity that
is $200$ times the critical density of the Universe at a given redshift.
Accretion/ejection from galaxies is defined based on whether particles are considered
part of the ISM. In turn, particles are considered part of the ISM
if they pass the \eagle star formation threshold based on density, temperature and metallicity
\cite[capturing the transition from the warm, atomic to the cold, molecular gas phase, as modelled by][]{Schaye04},
or that otherwise have total hydrogen number density $n_{\mathrm{H}} > 0.01 \, \mathrm{cm^{-3}}$ and are within
$0.5 \, \mathrm{dex}$ of the temperature floor corresponding
to the equation of state imposed on the unresolved ISM \cite[][]{Schaye08,Mitchell20a},
which corresponds roughly to the warm, atomic phase \cite[e.g.][]{Rahmati13},
above the density threshold for effective self-shielding from the far-ultraviolet radiation field.
With this definition, star-forming gas generally makes up the majority of the ISM in \eagle, with the additional
  non-star-forming selection picking out additional neutral gas in the outskirts.
All of these measurements are described and motivated fully in \cite{Mitchell20b, Mitchell20a}, including
details of (sub)halo definitions and merger trees.

\section{Results}
\label{result_sec}

\subsection{Baryonic mass budgets}
\label{mass_budget_sec}

\begin{figure}
\includegraphics[width=20pc]{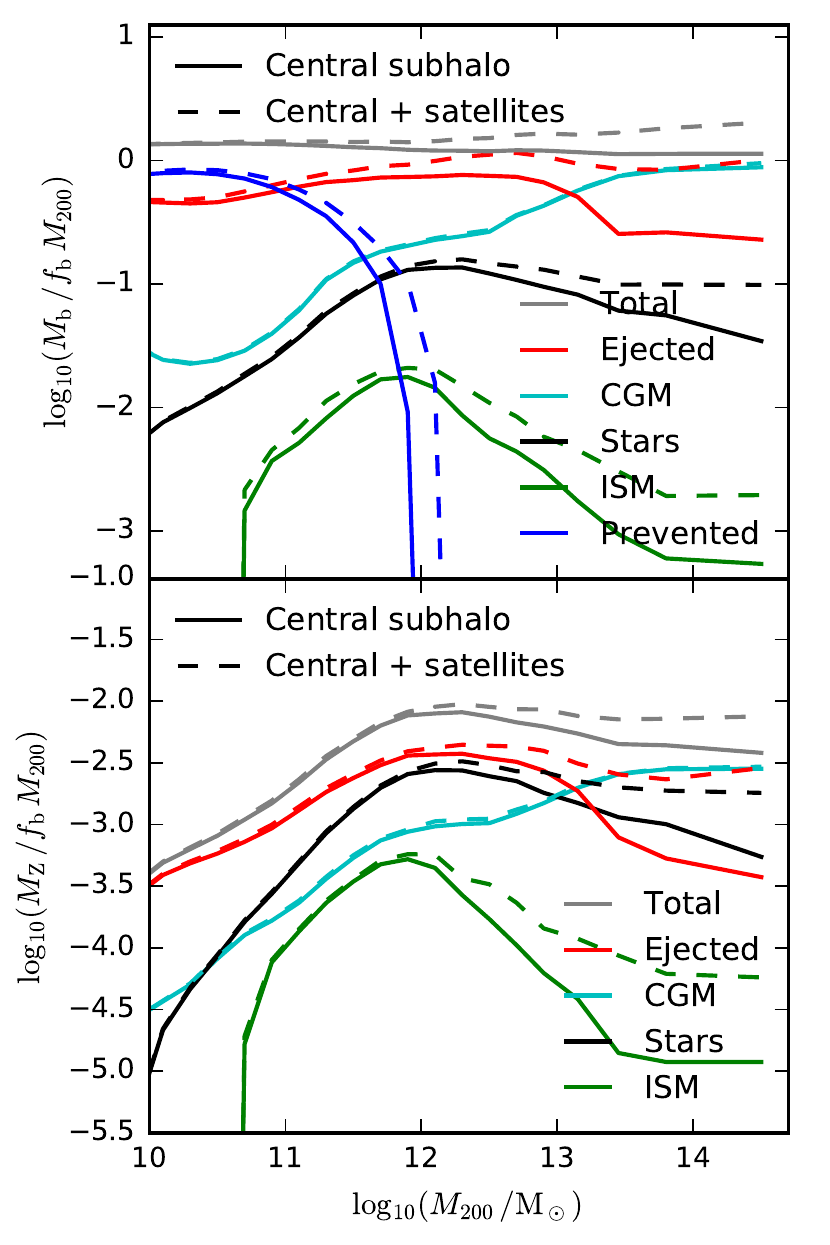}
\caption{The median total baryonic mass (top panel) and metal mass (bottom
panel) associated with the haloes of central galaxies at $z=0$, 
normalised by $\frac{\Omega_{\mathrm{b}}}{\Omega_{\mathrm{m}}} \, M_{200}$
and plotted as a function of halo mass ($M_{200}$).
Line colours indicate the mass in different components,
including the CGM (cyan), ISM (green), stars (black), gas
that has been ejected beyond $R_{200}$ (red), and gas
that we estimate has been prevented from being accreted
due to feedback effects (blue).
Grey lines show the total mass, adding together each of these components.
Solid lines show masses associated with the central subhalo,
whereas dashed lines also include the masses associated
with satellite subhaloes.
For $10^{11} < M_{200} \, / \mathrm{M_\odot} < 10^{13}$, most of the baryons that have ever been accreted
onto haloes have since been ejected and reside outside $R_{200}$ by $z=0$.
Preventative feedback is important for $M_{200} < 10^{12} \, \mathrm{M_\odot}$.
About half of the metals produced by stellar evolution are then ejected
beyond $R_{200}$, apart from in very massive haloes. 
}
\label{z0_mass_dep}
\end{figure}

Fig.~\ref{z0_mass_dep} presents mass budgets for baryons that are
associated with haloes at $z=0$, including the budget for all baryons
(top panel) and only metals (bottom panel). Baryons are
included in the budget either if they are within $R_{200}$ at $z=0$, divided
between stars (black lines), the ISM (green lines) and the CGM (cyan lines),
or have otherwise been ejected beyond the virial radius of a progenitor of the final
halo at any point in the past (red lines). We define the CGM
in this study simply as any gas that is within $R_{200}$, but
is not within the ISM as described in Section~\ref{measure_sec}. 
In addition, we also include baryons
that were expected to have been accreted onto the halo (given
the cumulative\footnote{By ``cumulative mass'', we are referring
  to the time-integrated mass of all dark matter (or gas) that has ever been
  uniquely accreted onto any progenitor of the subhalo in question. This
  does not double-count particles that are accreted more than
  once.}
mass of first-time dark matter accretion),
but have been prevented from doing so by feedback effects
\cite[blue line, top panel; see][for proof that feedback is responsible for this effect in \eagle]{Wright20}.
The mass in the ``prevented'' component is estimated via

\begin{equation}
M_{\mathrm{prev}} = \mathrm{max}\left( \frac{\Omega_{\mathrm{b}}}{\Omega_{\mathrm{m}}-\Omega_{\mathrm{b}}} \, M_{\mathrm{DM}}^{\mathrm{1st}} - M_{\mathrm{gas}}^{\mathrm{1st}} \, , \, 0\right),
\label{prev_eq}
\end{equation}

\noindent where $M_{\mathrm{gas}}^{\mathrm{1st}}$ is 
the cumulative mass off all baryonic particles that have ever
been accreted onto progenitors of the final subhalo for the first time
(i.e. recycled accretion is not double counted), 
and $M_{\mathrm{DM}}^{\mathrm{1st}}$ is the
corresponding cumulative mass of all first-time dark matter accretion. 
Here, ``first-time'' accretion refers to matter that
  has only been accreted once onto \emph{any} subhalo in the simulation,
  and ``recycled'' accretion refers to matter that has been accreted
  more than once.
Eqn.~\ref{prev_eq} assumes that without any preventative feedback effects,
first-time accretion of gas would exactly trace that of dark matter
(this is true at the $\approx 10 \, \%$ level
  in \eagle, comparing to simulation variants where feedback
  processes and/or radiative cooling have been disabled: Wright R.,
  private communication).
If first-time gas accretion does exactly trace (or exceed)
first-time dark matter accretion, then $M_{\mathrm{prev}} = 0$.

Masses are normalised by the expected baryonic mass
within $R_{200}$ (i.e., $f_{\mathrm{b}} M_{200} \equiv \frac{\Omega_{\mathrm{b}}}{\Omega_{\mathrm{m}}} \, M_{200}$),
assuming accretion/ejection of gas exactly traces that of dark 
matter.
Solid and dashed lines show baryonic mass budgets that exclude and
include respectively the mass associated with satellite subhaloes.
If satellites are included (dashed lines), then the gas that has been ejected from
(or prevented from being accreted onto) any progenitor of the
current satellite subhalo is included.
Note that the $M_{200}$ normalization in the denominator is the
same for both dashed and solid lines, and always includes all mass
within $R_{200}$.

Finally, we also show the sum of all the plotted baryonic components (including the prevented component, $M_{\mathrm{prev}}$)
as the ``total'' mass (grey lines). In general, the total mass always at least slightly exceeds
the basic expectation set by $f_{\mathrm{b}} M_{200}$. This is primarily because
$M_{200}$ by definition does not any include any gas that is ejected
beyond $R_{200}$ after having been accreted. We note also that 
if we include satellites, then the total mass (dashed grey line) increases
weakly but monotonically with increasing halo mass, at least for $M_{200} \gtrapprox 10^{12} \, \mathrm{M_\odot}$. 
This is qualitatively consistent with our findings in \cite{Mitchell20a}, where we found that in contrast to
preventative feedback, massive haloes actually accrete slightly more than their expected share of baryons,
with the excess growing fractionally with increasing halo mass. This is presumably connected to 
enhanced radiative cooling rates associated with metal enriched large-scale outflows.
Note however that the baryon fraction within $R_{200}$ (so excluding ejected gas) is always equal or less than the
cosmic baryon fraction.

Splitting into the various components, the ratio of galaxy stellar mass to halo mass peaks at the characteristic
halo mass $\approx 10^{12} \, \mathrm{M_\odot}$. The strength of this peak depends sensitively on whether satellites
are included in the stellar mass (as they are for the halo mass by convention), and also whether stellar
mass is defined within a spatial aperture to exclude the stellar halo 
\cite[see figure 1 and associated discussion in][]{Mitchell21}. Note that no spatial apertures are applied
for the stellar masses in Fig.~\ref{z0_mass_dep} (and in other figures).
If satellites are included, then the ratio of $M_\star / M_{200}$ depends only very weakly on halo mass for 
$M_{200} > 10^{12} \, \mathrm{M_\odot}$.

In contrast to the stellar mass, the mass in the CGM is always dominated by the central subhalo\footnote{
For central subhaloes we define the CGM as non-ISM gas that is within $R_{200}$, and that
  is not considered bound to a satellite by \subfind. For satellite subhaloes, the CGM is any non-ISM
gas that is considered bound to that satellite by \subfind.}. Note however
that this component is the most sensitive to the algorithm used to assign subhalo membership to particles.
If we only consider the central subhalo, the CGM is the largest single component for 
$M_{200} \gtrapprox 10^{13} \, \mathrm{M_\odot}$, and is in any case always the largest single contributor
to the baryonic mass within $R_{200}$. 

Generally speaking, much of the baryonic mass that has been accreted onto haloes resides outside $R_{200}$ at $z=0$,
even in the most massive galaxy clusters simulated with $M_{200} \sim 10^{14} \, \mathrm{M_\odot}$. Independent of
whether the gas ejected from satellites (and their progenitors) is included, the ejected gas reservoir
is the largest contributor over $10^{11} < M_{200} \, / \mathrm{M_\odot} < 10^{13}$, and is approximately
equal to the CGM mass in more massive haloes if satellites are included. This is contrary
to the notion that massive galaxy clusters retain\footnote{By ``retain'', we are
referring here to all of the baryons that have ever been part of the cluster halo and its progenitors,
and not to the more conventional consideration, which is simply a comparison of the current
baryonic mass fraction within the halo, relative to the universal fraction.}
all of their baryons, though note that the mass of
baryons within $R_{200}$ is still consistent with the universal baryon fraction in \eagle, see, e.g., \cite{Mitchell18}.

For halo masses $< 10^{11} \, \mathrm{M_\odot}$, the mass of gas
that we estimate is prevented from being accreted onto the halo 
exceeds that of any of the components that have been accreted. 
For halo masses $\gtrapprox 10^{12} \, \mathrm{M_\odot}$,
our definition of ``prevented'' mass yields zero, as in this range 
haloes actually accrete more than the expected number
of baryons in \eagle \cite[given the universal baryon fraction, see][]{Mitchell20a}.
Overall, we find the prevented component is comparable to the predictions
  from the zoom-in cosmological simulations presented by \cite{Christensen16},
  as inferred from their figure 7.

Finally, the ISM is always subdominant to the other components (at least at low redshift), and the median
actually drops to zero for $M_{200} \lessapprox 10^{11} \, \mathrm{M_\odot}$, due to a combination
of limited numerical resolution and stochastic feedback modelling. The $16^{\mathrm{th}}$ to $84^{\mathrm{th}}$ percentile scatter around
  the median at fixed halo mass (not shown for clarity of presentation) is notably larger for
  the ISM than for other components, reaching a local maximum of more than one dex at
  $M_{200} \approx 10^{12.5} \, \mathrm{M_\odot}$, and also increases at low halo mass due to
  sampling issues.

Given the model dependence of the results shown in Fig.~\ref{z0_mass_dep}, it is natural to
  question how this compares to predictions from other cosmological simulations, which implement
  subgrid feedback processes in different ways.
  While analysis of the ejected and prevented components shown here is not generally considered
  in studies of other cosmological simulations, some studies have considered the mass budgets associated
  with stars, the ISM, and the CGM. For example, \cite{Appleby21} present mass budgets as a function
  of stellar mass from the SIMBA cosmological simulations \cite[][]{Dave19}.
  At a mass scale corresponding to $M_{200} \sim 10^{12} \, \mathrm{M_\odot}$, SIMBA predicts
  that the total baryon content within a friends-of-friends group is $\approx 30 \%$ of the
  universal value, with $\approx 60 \, \%$ of those baryons in stars, and
  $\approx 30 \, \%$ in the CGM. At the same mass scale, \eagle predicts a slightly higher
  values for the baryon fraction within $R_{200}$ ($\approx 40 \, \%$), and a comparable balance of CGM, ISM
  and stars. In turn, the Illustris-TNG-100 simulation \cite[][]{Pillepich19} predicts gas fractions within $R_{200}$  that
  are more than twice the value predicted by \eagle \cite[][]{Davies20}, highlighting that the
  results shown in Fig.~\ref{z0_mass_dep} are indeed expected to be model dependent.

  In high-mass cluster haloes ($\sim 10^{14} \, \mathrm{M_\odot}$), SIMBA predicts that the
  baryon content is still lower than the universal value, at $70 \, \%$, where
  as \eagle predicts a value consistent with the Universal value at this mass scale
  within $R_{200}$. \cite{VanDaalen20} compare \eagle with other cosmological
  simulations at this same mass scale (their figure 16), finding that some simulations
  are consistent with the universal value (\eagle, Horizon-AGN, Illustris-TNG),
  but that other simulations (Illustris, and variants of the Cosmo-OWLS and BAHAMAS
  simulations) predict lower baryon fractions, again highlighting the model dependence.
  Finally, \cite{Lim21} compare the gas fractions within $R_{500}$ between the Magneticum simulation
  \cite[][]{Dolag16} with \eagle and Illustris-TNG-300 over a range of mass scales,
  finding notably that \eagle predicts values that are up to $\approx 50 \, \%$ larger
  than the other two simulations for $M_{500} \sim 10^{13} \, \mathrm{M_\odot}$.

Considering instead the mass in metals (bottom panel of Fig.~\ref{z0_mass_dep}), the total metal mass (grey lines) generally
traces the total baryonic mass in stars (black lines, top panel), since metal production is approximately proportional
to the star formation rate. Metals are first transferred back into the ISM before being
locked into later generations of stars, and can be ejected from the ISM in the intervening time.
For $M_{200} \lessapprox 10^{13} \, \mathrm{M_\odot}$ most of the metals that have been ejected from
galaxies have also ejected from their associated haloes. Most of the ejected metals are in the
CGM for more massive haloes, though as with total baryonic mass, even in galaxy clusters a comparable fraction of
the metals are ejected from haloes once satellite subhaloes (and their progenitors) are accounted for.

  \cite{Appleby21} present the relative mass in metals between stars, CGM and ISM within
  the SIMBA simulation. For $M_{200} \sim 10^{12} \, \mathrm{M_\odot}$, they
  find that $\approx 70 \, \%$ of the metals within haloes are in stars,
  which is close to the value of $\approx 60 \, \%$ we find for \eagle.
  For $M_{200} \sim 10^{14} \, \mathrm{M_\odot}$, this fraction drops
  is $\approx 50 \, \%$ in SIMBA, in contrast to \eagle where the corresponding
  fraction is $25 \, \%$ (with most of the metals being
  present in the hot CGM).

\begin{figure}
\includegraphics[width=20pc]{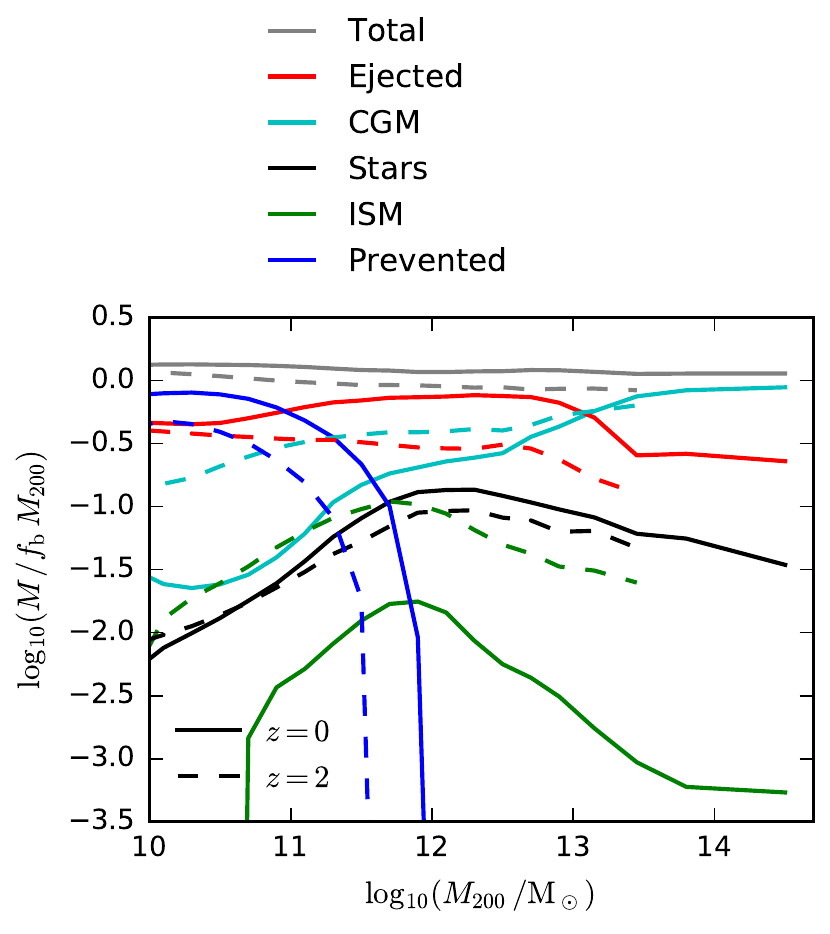}
\caption{A comparison of the median total baryonic mass in different components,
comparing haloes at $z=0$ (solid lines) with haloes at $z=2$.
Only the mass associated with central subhaloes is included.
At $z=0$, the plotted quantities are the same as the solid
lines shown in the top panel of \protect Fig.~\ref{z0_mass_dep}.
Feedback processes have ejected/prevented fewer baryons
from residing within $R_{200}$ at $z=2$, compared to at $z=0$.
}
\label{bar_mass_z2_comp}
\end{figure}

Fig.~\ref{bar_mass_z2_comp} compares the total baryonic mass budgets at $z=0$ with those
at $z=2$. As is to be expected given the higher specific star formation rates, 
the ISM represents a much larger mass fraction at $z=2$
compared to at $z=0$. 
This is also true for the CGM (at least for $M_{200} < 10^{12.5} \, \mathrm{M_\odot}$),
and in general the baryonic mass within $R_{200}$ is higher at $z=2$ than at $z=0$.
At fixed halo mass, feedback effects have prevented less gas from being accreted
at $z=2$ than at $z=0$, and also have ejected less gas outside $R_{200}$.
The halo mass scale over which preventative feedback is important
also increases with decreasing redshift.
For the ejected gas component located outside $R_{200}$, the low
efficiency of halo-scale gas recycling in \eagle \cite[][]{Mitchell20a} 
means that the mass in this component builds steadily over time, mirroring
the overall cosmic history of star formation in the simulation.

\subsection{How much of the CGM and the gas ejected from haloes has been processed through the ISM?}

\begin{figure}
\includegraphics[width=20pc]{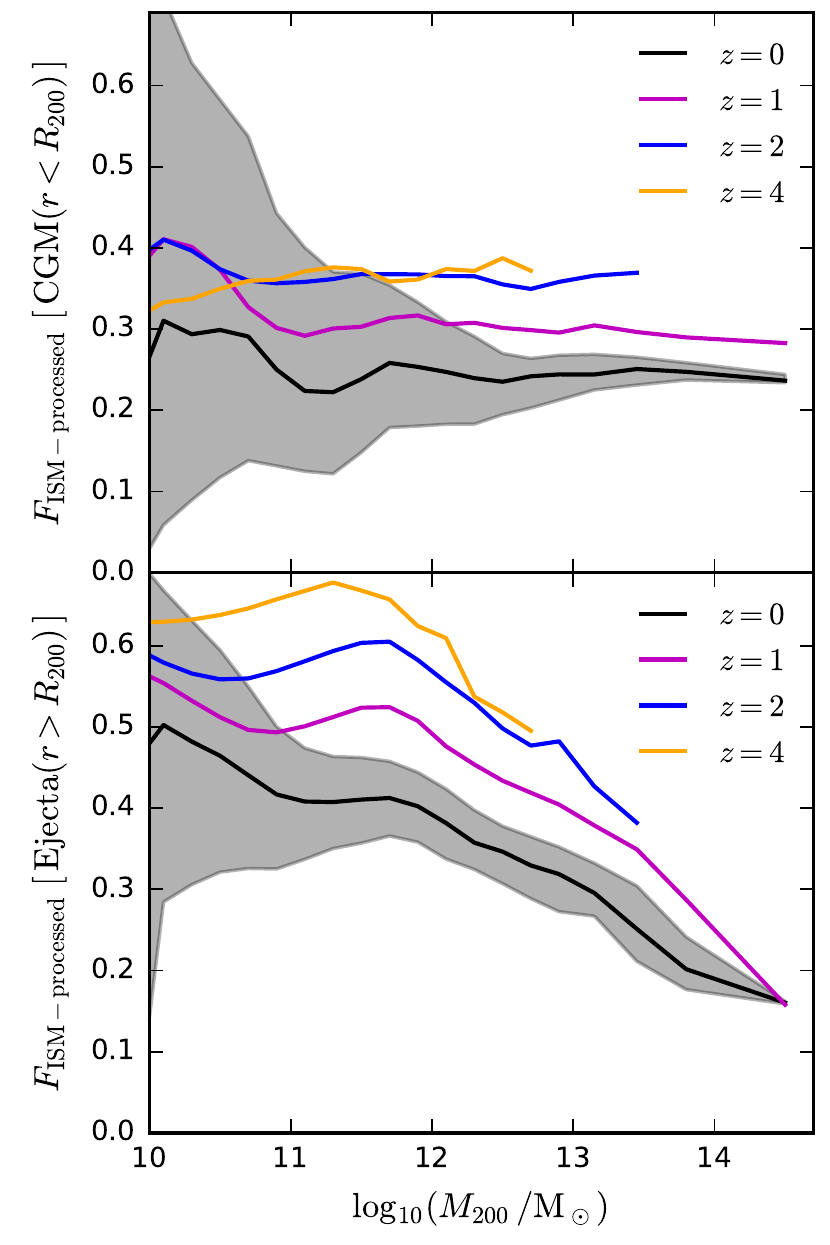}
\caption{The median mass fraction of gas that has been passed through the ISM of a galaxy in the past (``ISM-processed''),
plotted as a function of halo mass.
The top panel shows this fraction for gas in the CGM (with $r<R_{200}$), and the bottom panel
shows the corresponding fraction for gas in the ejected gas reservoir ($r > R_{200}$).
Different line colours correspond to different redshifts as labelled.
At $z=0$, we also show the $16$ to $84^{\mathrm{th}}$ percentiles of the distribution, as indicated by the shaded region.
Most of the gas in the CGM has not passed through the ISM in the past.
The same is generally true for the ejected gas reservoir outside
$R_{200}$, apart from for low-mass haloes.
}
\label{fwind_mhalo}
\end{figure}

Fig.~\ref{fwind_mhalo} shows what fraction of the mass in the CGM (top panel) 
and the ejected gas reservoir outside $R_{200}$ 
(bottom panel) has passed through the ISM of a galaxy in the past (``ISM-processed'').
This is computed by tracking gas particles forwards
  in time through the recorded snapshots of the simulation, and flagging
  whether the particle ever passed the ISM selection criteria described
  in Section~\ref{measure_sec}.
The ISM-processed fraction is nearly independent of halo mass for gas in the CGM,
though a slight anti-correlation is apparent at lower redshifts. 
The fraction is $\approx 25 \, \%$ at $z=0$, increasing with redshift
up to an apparent maximum value of $\approx 40 \, \%$ at $z=2$ and above (the fraction
eventually decreases again for even higher redshifts that are not plotted).
Across all masses and redshifts, the majority of the mass in the CGM has not been 
processed through the ISM at an earlier stage. This picture is consistent with 
recent results from another set of hydrodynamical simulations \cite[][]{Hafen19},
and from empirical modelling of observations \cite[][]{Afruni21}.
The scatter around the median (shown for $z=0$) does change significantly
  with halo mass, decreasing strongly with increasing halo mass. We expect
  this is driven in part by the hierarchical assembly of haloes (with more
  massive haloes representing the aggregate average over many smaller haloes),
  and in part by numerical resolution in conjunction with
  the efficient stellar feedback in low-mass haloes, as the ISM content
  of low-mass haloes ($M_{200} < 10^{11} \, \mathrm{M_\odot}$)
  and their satellites is heavily under-sampled (see, e.g., Fig.~\ref{z0_mass_dep}).

For the ejected gas reservoir outside $R_{200}$ (bottom panel) the ISM-processed
fraction is larger than in the CGM for low-mass haloes, but depends negatively
on halo mass for $M_{200} > 10^{12} \, \mathrm{M_\odot}$, and drops to $\approx 20 \, \%$
in massive galaxy clusters. The fraction
increases monotonically with increasing redshift.
As the Universe expands and densities and accretion
rates decrease, an ambient CGM can increasingly develop, meaning that feedback-driven outflows
increasingly have to sweep up more and more CGM material (that the top panel shows typically has 
not been processed
through the ISM) before ultimately escaping the halo. 

\subsubsection{Temperature and radius dependence}

\begin{figure}
\includegraphics[width=20pc]{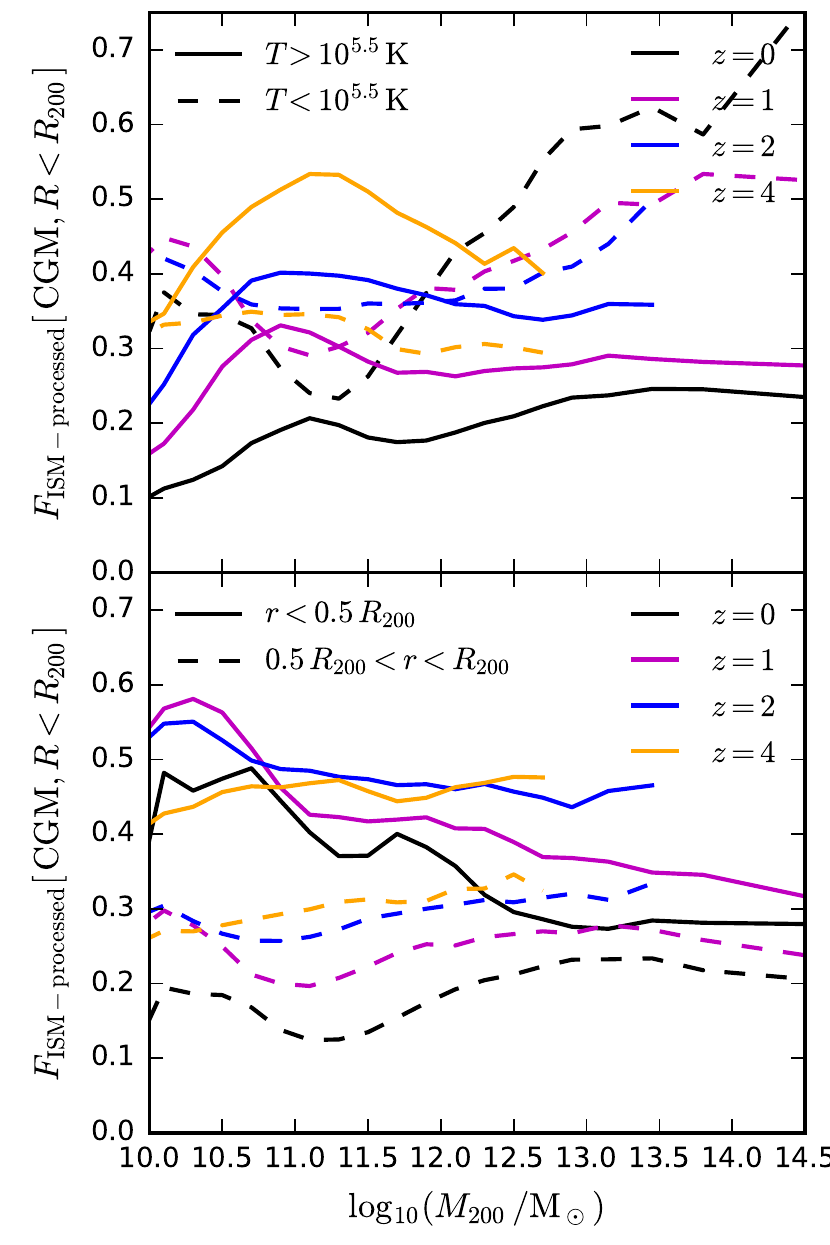}
\caption{Median ISM-processed mass fractions for circum-galactic gas within haloes, in this case divided into two bins of either temperature (top panel)
or radius (bottom panel).
In the top panel solid lines correspond to hot gas ($T>10^{5.5} \, \mathrm{K}$), and dashed lines refer to cool/warm gas ($T < 10^{5.5} \, \mathrm{K}$).
In the bottom panel solid lines correspond to the inner CGM ($R<0.5 \, R_{200}$), and dashed lines refer to the outer CGM ($0.5 R_{200} < r < R_{200}$).
Different line colours correspond to different redshifts, as labelled.
For the cool/warm gas component, most of the gas has passed through the ISM in massive haloes, in contrast to to the hot gas component.
The ISM-processed fraction is significantly higher in the inner CGM, compared to the outer CGM.
}
\label{fwind_mhalo_hotcold}
\end{figure}

Fig.~\ref{fwind_mhalo_hotcold} shows the mass fraction of the CGM that has been processed
earlier through the ISM, splitting now either by gas temperature between hot 
($T > 10^{5.5} \, \mathrm{K}$, solid lines) and cool/warm
($T < 10^{5.5} \, \mathrm{K}$, dashed lines) phases (top panel), or by radius between
the inner ($r < 0.5 R_{200}$, solid lines) and outer ($r > 0.5 R_{200}$, dashed lines) regions of the halo (bottom panel).
Focussing first on the top panel, we see that for the cooler gas selection,
the ISM-processed fraction does not depend monotonically on halo mass,
reaching an apparent minimum value at $M_{200} \sim 10^{11} \, \mathrm{M_\odot}$.
For $M_{200} \gtrsim 10^{13} \, \mathrm{M_\odot}$, the ISM-processed
fraction for the cooler gas selection can actually reach (or even exceed) $50 \, \%$,
which is markedly different from the situation for hot gas (solid lines), or for the total
CGM mass budget (Fig.~\ref{fwind_mhalo}). 
For the hotter gas selection, the dependence on halo mass is comparatively much weaker, but
is notably more dependent on redshift at fixed halo mass.

Focussing instead on the bottom panel of Fig.~\ref{fwind_mhalo_hotcold}, we 
find, unsurprisingly, that gas in the
inner CGM is more likely to have passed through the ISM in the past ($\approx 40 \, \%$),
when compared to the outer CGM ($\approx 20 \, \%$). The ISM-processed fraction
increases slightly with redshift up to $z=2$ for both the inner and outer CGM, and trends
with halo mass are modest.

\subsubsection{Origin of the ISM-processed circum-galactic gas}

\begin{figure}
\includegraphics[width=20pc]{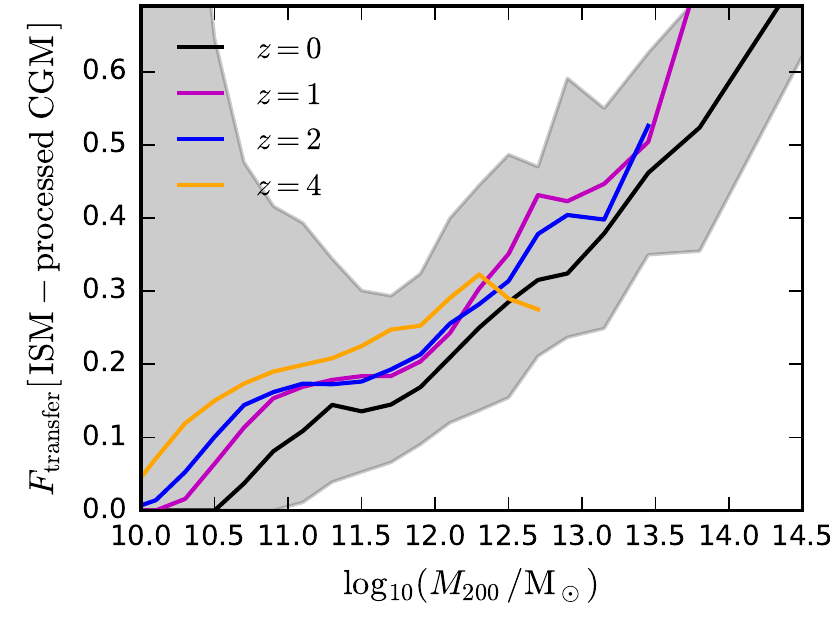}
\caption{For the subset of circum-galactic gas that has passed through the ISM in
the past (``ISM-processed''), the median mass fraction of that gas that
comes from (the progenitor of) a non-progenitor galaxy (i.e. ``transferred'' 
from satellites) is plotted as a function of halo mass.
Different line colours correspond to different redshifts, as labelled.
At $z=0$, we also show the $16$ to $84^{\mathrm{th}}$ percentiles of the distribution, as indicated by the shaded region.
ISM-processed gas originates almost exclusively from progenitors of
the current central galaxy in low-mass haloes, but gas ejected/stripped
from non-progenitor galaxies (generally satellites) is increasingly important for more massive haloes.
}
\label{fgaltran_mhalo}
\end{figure}

Fig.~\ref{fgaltran_mhalo} shows how much of the ISM-processed CGM was last ejected
from the ISM of non-progenitor versus progenitor galaxies (i.e., from satellites
versus from the central, at least in most cases). This corresponds to the definition
of ``recycled'' versus ``transferred'' gas accretion, as discussed in \cite{Mitchell20b}
\cite[see also][]{AnglesAlcazar17,Grand19, Hafen19}. The former generally
refers to gas ejection from the ISM of a central galaxy via feedback processes, whereas
the latter will be triggered by a mix of feedback plus stripping via ram pressure, gravitational tides, etc.

The plotted ``transfer'' fraction depends positively on halo mass. Only $\approx 20 \, \%$
of the ISM-processed CGM originates from the progenitors of satellites
for $M_{200} \sim 10^{12} \, \mathrm{M_\odot}$, but this fraction
increases to $\approx 50 \, \%$ for $M_{200} \sim 10^{13.5} \, \mathrm{M_\odot}$.
While not shown for conciseness, we also find that this non-progenitor
fraction depends on radius. For $M_{200}(z=0) \sim 10^{12} \, \mathrm{M_\odot}$,
$F^{\mathrm{ISM-processed}}_{\mathrm{transfer}}$ is only $\approx 10 \, \%$ for $r<0.5 \, R_{200}$,
but reaches $\approx 40 \, \%$ for $r > 0.5 \, R_{200}$. No significant trend
with gas temperature is apparent, however.

\subsection{Tracking the evolution of the halo baryonic mass budget}

\begin{figure*}
\begin{center}
\includegraphics[width=40pc]{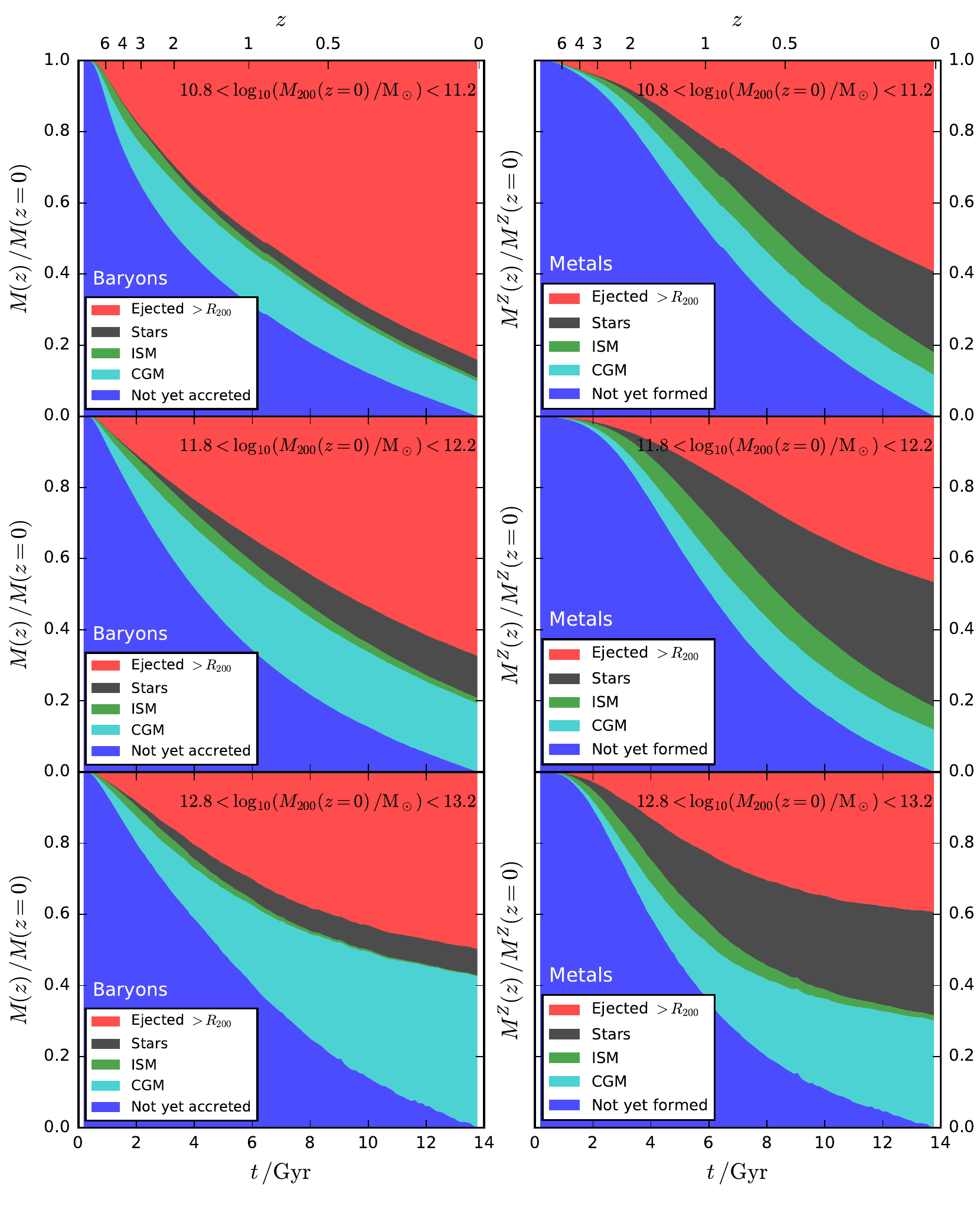}
\caption{
The total baryonic mass (left panels) and metal mass (right panels) fractions
associated with different components, plotted as a function
of cosmic time. Central subhaloes are selected by halo mass at $z=0$, and
tracked backwards in time. For each sample of haloes,
all baryons that are ever accreted onto progenitor subhaloes are included.
Each row corresponds to a different halo mass bin, as labelled. 
The left column shows mass fractions for
gas that has not yet been accreted (blue), the CGM (cyan), ISM (green),
stars (black), and gas that has been ejected beyond $R_{200}$.
The right column shows analogous results for metals, where
``not yet accreted'' is replaced by ``not yet formed''.
}
\label{halo_centric}
\end{center}
\end{figure*}

Fig.~\ref{halo_centric} presents a visual overview of the history
of baryonic accretion/ejection onto haloes in the \eagle simulation.
To make this figure, we select central subhaloes by halo mass
at $z=0$, and track their progenitors backwards in time.
Starting from high redshift, all of the baryons that will ever be accreted 
onto the subhalo progenitors of the selected
galaxies are included, plotted as mass fractions that are split into different
components. For the left panels, these components include gas that has not yet
been accreted onto a progenitor subhalo (blue), gas that is part of the CGM (cyan),
ISM (green), stars (black), and gas that has been ejected beyond $R_{200}$
(red). For the right-side panels, mass fractions are instead
defined by the metal mass in each component, and correspondingly the
first component (blue) refers to metals that have not yet formed.
Note that we do not include a ``prevented'' gas component in this
figure (i.e., we only include gas that is actually accreted at some point
during the evolution of the haloes).
Note also that the mass fractions within haloes at a particular time can be obtained
by ignoring the blue (``not yet accreted/formed'') and red (``ejected $>R_{200}$'')
shaded regions, and renormalizing the three remaining components so that
their sum equals unity.

Focussing first on the total baryonic mass fractions (left panels),
we see that (as in earlier figures) 
most of the gas that has been accreted onto haloes has
since been ejected and resides beyond $R_{200}$ by $z=0$.
The mass fraction in the stellar and ejected components
increases monotonically with cosmic time, and the ISM
  mass fraction peaks at high redshift. For the CGM,
  the mass fraction peaks at $z \approx 2$ for $M_{200}(z=0) \sim 10^{11} \, \mathrm{M_\odot}$,
  grows with time until $z=1$ for $M_{200}(z=0) \sim 10^{12} \, \mathrm{M_\odot}$ (and is steady at later times),
  and grows monotonically with cosmic time for $M_{200}(z=0) \sim 10^{13} \, \mathrm{M_\odot}$.
which peaks at high redshift.
It is apparent that lower mass haloes first accrete their associated baryons
earlier: $50 \, \%$ of the baryons that will ever accrete do so by $z=1.9, 1.5$ and $1.3$
respectively for $M_{200}(z=0) \sim 10^{11} \, \mathrm{M_\odot}, 10^{12} \, \mathrm{M_\odot}$, 
and $10^{13} \, \mathrm{M_\odot}$.

To understand these trends in \eagle, we can refer back to the earlier analyses
  of inflow and outflow rates presented in \cite{Mitchell20b, Mitchell20a, Mitchell21}.
  For example, one might suspect that the differences in mass fractions in Fig.~\ref{halo_centric}
  (and Fig.~\ref{z0_mass_dep}) between different halo mass ranges would be caused by a scale-dependence of
  halo-scale gas outflows, in the sense that higher-mass haloes eject comparatively
  less gas (per unit halo mass) than lower-mass haloes. This is in fact not the
  case at least to leading order
  \cite[figure 1, bottom-right panel,][the halo-scale outflow per unit halo mass is nearly independent of halo mass]{Mitchell20a}.
  Rather, the main halo-mass dependence in Fig.~\ref{halo_centric}
  is driven primarily by halo-scale gas recycling. 
  The characteristic recycling timescale for the return of ejected gas
  scales decreases with halo mass \cite[figure 7, top panel,][]{Mitchell20b}, such that the vast majority of gas
  ejected from low-mass haloes ($M_{200} \ll 10^{12} \, \mathrm{M_\odot}$) never returns,
  whereas the gas ejected from high-mass haloes ($M_{200} \gg 10^{12} \, \mathrm{M_\odot}$)
  is able to return in less than a Hubble time. 
  Incidentally, this is qualitatively consistent with
  the inferred gas return time dependence inferred by fitting a semi-analytic galaxy
  formation model to various observational constraints related to the stellar content
  of galaxies \cite[][]{Henriques13}, and can be rationalized physically simply by
  noting that more massive haloes require more energy per unit star formation (or
  per unit supermassive black hole growth) for gas to be able to
  permanently escape the halo.

To understand why more massive haloes first accrete their associated baryons later,
  we can first consider the weak anti-correlation between halo concentration and halo
  mass, which is known to reflect in turn a weak dependence of halo formation time on
  the final halo mass, in the sense that more massive $z=0$ haloes assembled their
  mass slightly later \cite[e.g.,][]{Bullock01, Wechsler02, Correa15b}. In addition, we can also expect
  that preventative feedback (which is only effective in \eagle for $M_{200} < 10^{12} \, \mathrm{M_\odot}$, Fig.~\ref{z0_mass_dep})
  may also shape this further, in the sense that preventative feedback is seemingly
  more significant at $z=0$ than at $z=2$ (Fig.~\ref{bar_mass_z2_comp}), with the
  consequence that the baryons that are able to be accreted onto low-mass haloes
  are accreted earlier, preferentially.

Focussing instead on the corresponding evolution in metal mass
fractions (right panels), we see that
metal formation is substantially delayed with respect to first-time
gas accretion onto haloes, particularly in low-mass haloes.
As seen in earlier figures, haloes retain a higher fraction of their metals
than they do for total baryonic mass, meaning that gas ejected from the halo is
on average metal deficient compared to the baryons within $R_{200}$.
Yet still, almost half the metals have been ejected
outside $R_{200}$ by $z=0$ for $M_{200}(z=0) \sim 10^{12} \, \mathrm{M_\odot}$.
For $M_{200}(z=0) \sim 10^{12} \, \mathrm{M_\odot}$ most
of the remaining (non-halo-ejected) metals are in stars,
and only $12 \, \%$ are in the CGM (defined $<R_{200}$).

In a recent census of cosmic metals undertaken by \cite{Peroux20}, it
  is estimated that $50 \, \%$ of cosmic metals are in stars at $z=0$ \cite[see however also][their figures 5,9]{Peeples14}, and
  that at high-redshift almost all of
  the expected metals (given the observationally inferred star formation rate density)
  are accounted for by damped Lyman-$\alpha$ absorbers (DLAs) at $z\approx 4.5$,
  and by a combination of DLAs and lower column density absorbers at $z=3.5$.
  If the scenario presented by \eagle is
  quantitatively correct, this would imply that the observationally inferred low-redshift metal
  fraction locked in stars is over-estimated. This would also indicate
  (assuming in this case that both observational inference and \eagle
  are correct)
  that the majority of observed metal absorbers at high-redshift
  are spatially located outside the virial radius of haloes.

\subsubsection{Tracking all the baryons that will ever be accreted onto galaxies}

\begin{figure*}
\begin{center}
\includegraphics[width=40pc]{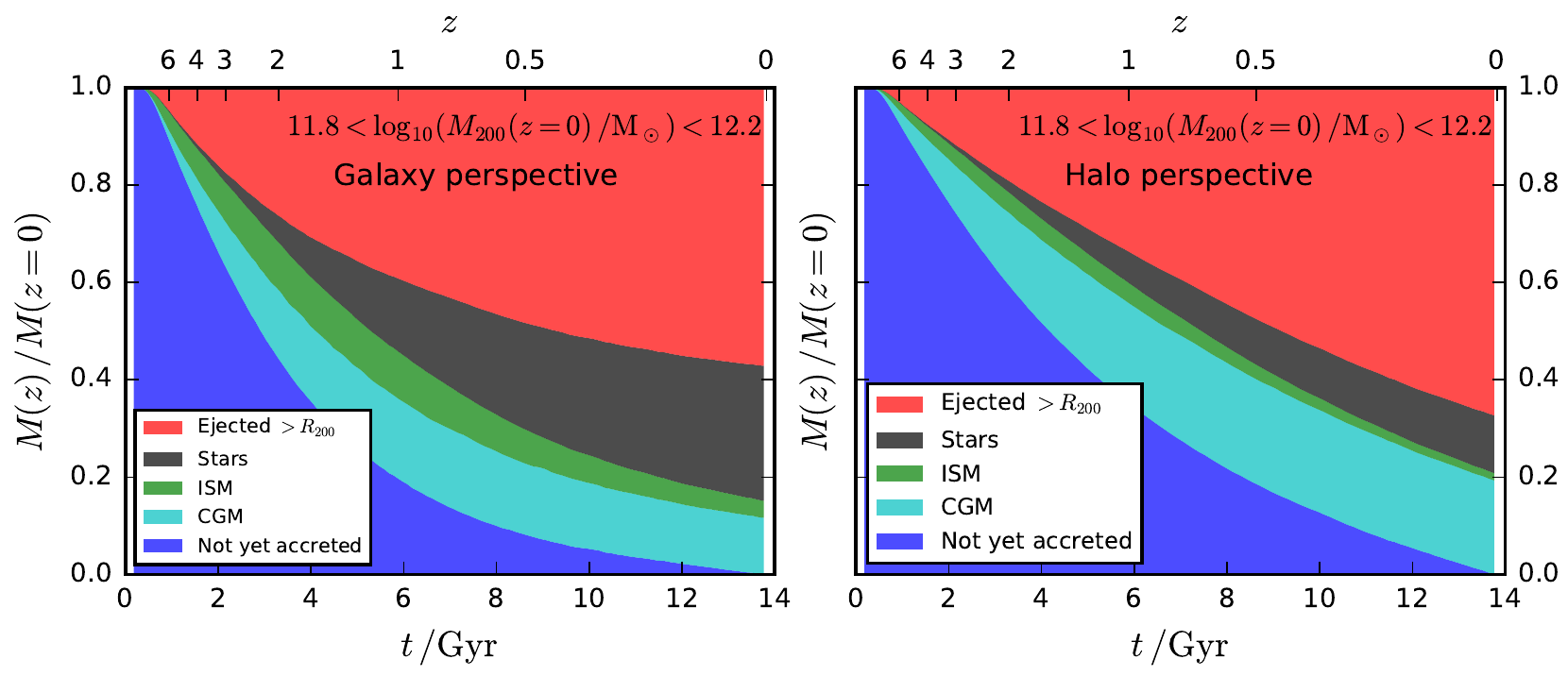}
\caption{
The time evolution of mass fractions for all the baryons that have ever been accreted onto
progenitor galaxies (left panel), compared to all those that have ever been accreted onto
progenitor haloes (right panel), 
for haloes selected to have $M_{200} \sim 10^{12} \, \mathrm{M_\odot}$ at $z=0$.
Mass fractions are shown for
gas that has not yet been accreted (blue), the CGM (cyan), ISM (green),
stars (black), and gas that has been ejected beyond $R_{200}$.
}
\label{gal_centric}
\end{center}
\end{figure*}

To supplement Fig.~\ref{halo_centric}, Fig.~\ref{gal_centric} shows the 
``galaxy-centric'' mass fractions for the subset of baryons that
have ever been accreted onto the ISM of a progenitor galaxy (rather
than all those that are accreted onto a progenitor halo at $R_{200}$).
The left panel in Fig.~\ref{gal_centric} show this galaxy-centric
view, otherwise repeating the format of Fig.~\ref{halo_centric}
for the $M_{200}(z=0) \sim 10^{12} \, \mathrm{M_\odot}$ mass bin.
The right panel in Fig.~\ref{gal_centric} repeats the halo-centric
information in Fig.~\ref{halo_centric} as a reference.
Note that the galaxy-centric perspective is equivalent to the halo-centric
perspective with the non-ISM-processed CGM and non-ISM processed ejected gas components removed.

While relatively fewer of the baryons that are accreted onto galaxies are ejected
beyond $R_{200}$ (compared to all of the baryons that are accreted onto haloes),
it remains the case that the majority are still in this ejected component by $z=0$
for $M_{200}(z=0) \sim 10^{12} \, \mathrm{M_\odot}$.
  As for other halo mass ranges (not shown), the same is true for lower-final-mass haloes, but for more
  massive haloes ($M_{200}(z=0) \sim 10^{13} \, \mathrm{M_\odot}$)
  the ejected mass fractions are similar for the baryon sets that
  have passed through galaxies compared to those that have passed through
  haloes, and this component makes
  up less than $50 \, \%$ of the total ($40 \, \%$ for $M_{200}(z=0) \sim 10^{13} \, \mathrm{M_\odot}$).

For $M_{200}(z=0) \sim 10^{12} \, \mathrm{M_\odot}$, and considering
  only the baryons that are retained within the halo, at $z=0$ most are in stars, a minority
are in the CGM, and only a few percent are in the ISM, despite gas
return to the ISM through stellar mass loss.
Irrespective of the final halo mass,
galaxies generally accrete their baryons slightly earlier than haloes,
half of the accretion has occurred by $z \approx 2.5$ for galaxies,
compared to $z \approx 1.5$ for haloes.
This is because haloes continue to accrete significant mass
at late times (partly due to pseudo-evolution effects associated
with the definition of the virial radius), whereas galaxy-scale gas
accretion becomes inefficient at late times \cite[see, e.g., ][]{Vandevoort11b,Mitchell20b}.

\section{Summary}
\label{summary_sec}

We have presented predictions for the mass budgets of gas and metals
around galaxies and haloes in the \eagle cosmological simulation, splitting between 
stars, the ISM, CGM, gas that has been ejected beyond $R_{200}$, and 
gas that was prevented from having ever been accreted in the first place.
We also quantify how much of the CGM has passed through the ISM
of a galaxy in the past, and whether such ``ISM-processed'' gas
originated from progenitors of the current galaxy.

We find that in general the majority of gas that has ever been accreted onto
haloes has since been ejected, and resides beyond $R_{200}$ at
$z=0$ (Fig.~\ref{z0_mass_dep}, Fig.~\ref{halo_centric}).
Perhaps surprisingly, even though most massive simulated galaxy
clusters ($M_{200} \sim 10^{14} \, \mathrm{M_\odot}$) have baryon fractions close the universal
value, the mass in ejected gas is still $50 \, \%$ once the gas that was
ejected from progenitors of current satellites is accounted for (Fig.~\ref{z0_mass_dep}).
For the metals that are produced by stellar evolution, about half 
are generally predicted to reside outside $R_{200}$, with
the remainder either mostly in stars ($M_{200} \sim 10^{12} \, \mathrm{M_\odot}$)
or in the CGM/intra-cluster medium ($M_{200} \gtrsim 10^{13} \, \mathrm{M_\odot}$).

For haloes with mass $M_{200} < 10^{12} \, \mathrm{M_\odot}$, many
of the baryons that were expected to be accreted (given the total
halo mass) were prevented from doing so by feedback processes (Fig.~\ref{z0_mass_dep}).
This ``prevented'' component comprises about half of
the total baryon mass budget for $M_{200} < 10^{11} \, \mathrm{M_\odot}$,
and less for higher masses.
At $z=2$, the prevented mass (and also the ejected mass) represents
a smaller fraction of the total baryon budget than at $z=0$ (Fig.~\ref{bar_mass_z2_comp}),
as feedback has had less time to cumulatively shape the baryon
content in and around haloes.

We find that most of the mass in the CGM has not passed through the ISM of a progenitor galaxy, 
for all halo masses and redshifts (Fig.~\ref{fwind_mhalo}).
This is also true for the gas that has been ejected
beyond $R_{200}$, except for the CGM of low-mass haloes ($M_{200} < 10^{11} \, \mathrm{M_\odot}$),
where more than half of the ejected baryons have passed
through the ISM in the past.
The fraction of the CGM that has passed through
the ISM is higher in the inner CGM ($r < 0.5 \, R_{200}$)
than in the outer CGM $0.5 \, R_{200} > r > R_{200}$ (Fig.~\ref{fwind_mhalo_hotcold}).
In high-mass haloes ($M_{200} > 10^{12} \, \mathrm{M_\odot}$),
cool circum-galactic gas ($T < 10^{5.5} \, \mathrm{K}$) 
is significantly more likely to come from the ISM than hotter gas 
($T > 10^{5.5} \, \mathrm{K}$), but this trend weakens
(or even reverses) in lower mass haloes (Fig.~\ref{fwind_mhalo_hotcold}).
We also find that for low- and intermediate-mass haloes most of the CGM
that has passed through the ISM in the past originates from progenitors
of the current central galaxy
(Fig.~\ref{fgaltran_mhalo}), but that in very massive halos
($M_{200} > 10^{13} \, \mathrm{M_\odot}$)
most of the CGM that has passed through the ISM comes
from non-progenitor galaxies (typically the progenitors of current
satellites).

Finally, we present summary figures that provide a useful
pedagogical overview of how haloes (Fig.~\ref{halo_centric}) and 
galaxies (Fig.~\ref{gal_centric}) accrete (and eject) their baryons
over cosmic time. These figures highlight that lower-mass
haloes first accrete their baryons earlier than more massive
haloes, that metal formation is significantly delayed
with respect to the first baryonic accretion onto haloes
(Fig.~\ref{halo_centric}), and that galaxies actually
accrete their associated baryons (that will ever be
accreted by $z=0$) relatively earlier than haloes,
since haloes continue to accrete significant
material at late times while galaxy-scale
accretion shuts down (Fig.~\ref{gal_centric}).

These predictions are clearly model dependent, but
  basic comparisons of the integrated baryon content within
  haloes between different cosmological simulations
  (see discussion in Section~\ref{mass_budget_sec}) show
  encouraging qualitative agreement (yet still with some interesting
  quantitative differences), mirroring the good qualitative
  agreement (and quantitative agreement in some cases)
  we found when comparing \eagle to other
  simulations in terms of inflow and outflow rates
  \cite[][]{Mitchell20b, Mitchell20a}. Ultimately, robust observational
  tests will be needed to push the subject further, with
  promising progress being made for example in constraining
  the integrated baryon content of haloes via the
  Sunyaev-Zel'dovich effect \cite[e.g.][]{Hill18, Lim20, Tanimura20, Wu20, Pratt21, Vavagiakis21}.
  Observations of hydrogen and metal absorption
  naturally should also constrain the predictions from cosmological simulations.
  We close by noting that for simulators to fully harness
  the constraining power of such observations,
  our results reiterate that more mature modelling
  of the mixing of metals with pristine gas will
  presumably be needed, given   the \eagle prediction
  that most of the
  gas that is displaced from haloes has not
  been processed directly through the ISM of a galaxy.

\section*{Acknowledgements}

This work used the DiRAC@Durham facility managed by the Institute for
Computational Cosmology on behalf of the STFC DiRAC HPC Facility
(www.dirac.ac.uk). The equipment was funded by BEIS capital funding
via STFC capital grants ST/K00042X/1, ST/P002293/1, ST/R002371/1 and
ST/S002502/1, Durham University and STFC operations grant
ST/R000832/1. DiRAC is part of the National e-Infrastructure.

This work was supported by Vici grant 639.043.409 from
the Netherlands Organisation for Scientific Research (NWO).

\section*{Data availability}

The data underlying this article will be shared on reasonable request to the corresponding author.
Raw particle data and merger trees for the \eagle simulations have been publicly released \cite[][]{McAlpine16}.

\bibliographystyle{mn2e}
\bibliography{bibliography}

\label{lastpage}
\end{document}